# Predicting Community College Astronomy Performance through Logistic Regression

Zachary Richards
*Department of Earth & Physical Sciences, York College, City University of New York,
94-20 Guy R. Brewer Blvd, Jamaica, NY 11451, USA
Institute for STEM Education, Stony Brook University, 092 Life Sciences, Stony Brook, NY
11794-5233, USA*

Angela M. Kelly[1]
*Department of Physics & Astronomy and Institute for STEM Education,
Stony Brook University, 092 Life Sciences, Stony Brook, NY 11794-5233, USA*

The present study examined demographic and academic predictors of astronomy performance among a cohort of *N*=1909 community college students enrolled in astronomy courses in a large suburban community college during a four-year timeframe, 2015-2019. The theoretical framework was based upon a deconstructive approach for predicting community college performance, whereby students' academic pathways through higher education institutions are examined to understand their dynamic interaction with institutional integration and progress towards academic goals. Transcript data analysis was employed to elicit student demographics and longitudinal academic coursework and performance. A logistic regression model was generated to identify significant predictors of astronomy performance, which included mathematics achievement, enrollment in remedial mathematics, and enrollment in multiple astronomy courses. The results imply a greater focus on mathematics preparation and performance may mediate astronomy outcomes for community college students. Notably, demographic variables including ethnicity, socioeconomic status, gender, and age were not significant predictors of astronomy performance in the multivariable model, suggesting the course is a potential gateway for diversifying STEM access. Also, astronomy interest, as measured by enrollment in multiple astronomy courses was related to performance. Further implications for practice are discussed.

[1] angela.kelly@stonybrook.edu

# I. INTRODUCTION

Community college students are a largely untapped resource in meeting the need for a diversified, well-prepared science, technology, engineering, and mathematics (STEM) workforce in the United States. Astronomy is a course taken by many community college students to fulfill general natural science requirements. It is often considered a gateway science since it is a course that many community college students take [1], potentially fostering scientific literacy and interest in STEM. The present study examined the predictive value of academic and demographic variables with respect to the astronomy performance of community college students.

Research on the STEM performance of community college students has been a limited line of inquiry in science education research. This is notable since approximately 50% of all undergraduate students in the United States enroll at a community college at some point in their academic careers [2]. As of 2021, there were 6.8 million students in community colleges and approximately 65% of them were part-time students [3]. Since there is a high demand for STEM workers in the private sector, particularly for jobs requiring associate degrees [4], community college students are a diverse resource who may be well equipped to meet this need. Consequently, research on their STEM enrollment and performance is needed to maximize their STEM preparation and potential.

Astronomy was chosen as the disciplinary focus of this work due to the prevalence of enrollment in this course in higher education; one out of ten undergraduate students, or approximately 250,000 students, will take an astronomy course during college [5]. In 2011, there were approximate the same number of students taking astronomy in community colleges (51,000) as there were taking astronomy (52,000) in four-year institutions [1]. Astronomy is often taken by non-science majors to fulfill general education requirements for graduation and may be an undergraduate student's only experience with formal science coursework [6-8]. Consequently, astronomy may be formative in developing students' scientific literacy and inspiring students' interest in science [9].

It is worthwhile to examine factors that predict astronomy coursetaking and performance to maximize students' access to science and their overall success in community college. This rationale has prompted the following research question to identify target interventions: How do students' academic coursework and performance and demographic characteristics predict introductory community college astronomy performance?

The present study examined astronomy coursetaking and performance among a sample of $N$=1909 students over a three-year period in a community college in the Northeastern United States. Transcript analysis provides a road map of academic and demographic variables that may identify predictors of community college student outcomes, particularly in terms of academic prerequisites, key decision points, and the identification of potential interventions for specific groups of students [10]. Binary logistic regression was employed in this study since it has been shown to be an effective method of predicting dichotomous student outcomes in higher education research [11-17].

# II. LITERATURE REVIEW

## A. Community college context and STEM coursetaking

Researchers have documented several academic roadblocks related to STEM performance and persistence for community college students, particularly in mathematics. Community college STEM students have often been underprepared for higher education and require mathematics remediation at a higher rate (59%) when compared to four-year STEM students (23%) [15]. Mathematics coursetaking and grade point average (GPA) have been cited as more predictive of STEM persistence and attrition than demographic variables [14], although this has been contradicted in other studies that identified demographics as prevalent predictors of STEM outcomes [18-20]. Students who enter community college enrolling in a developmental mathematics course experience higher levels of STEM attrition, and mathematics enrollment and performance predict science outcomes [13,14,16]. Mathematics is often an academic gatekeeper in STEM and may deter students from pursuing a STEM major and or finishing their degree entirely [14,21].

### B. Astronomy in higher education

Mathematics has often been applied in introductory astronomy courses, and may have a relationship with student performance, although the mathematical rigor of an astronomy course has not been shown to improve conceptual understanding of astronomical concepts [22]. Approximately 33% of students enrolled in astronomy courses have only completed algebra [6], indicating potentially insufficient mathematical background to fully grasp some astronomy topics.

The mathematics involved in introductory astronomy typically includes basic algebra to solve kinematics, dynamics, and electromagnetic spectrum problems [23]. For example, when discussing motions of solar system bodies, it is common practice to discuss Kepler's laws of planetary motion – specifically, Kepler's third law, which states the further away a planet is from the Sun, the longer it will take to complete one orbit,

$$p^2 = a^3. \quad (1)$$

The ratio of the square of the period to the cube of the mean distance is constant for each planet. Here $p$ is the time to complete an orbit around the Sun in years and $a$ is the semi-major axis or distance from the Sun in astronomical units. To solve this equation, students need to use squares, cubes, both square and cubic roots, and in some cases, conversion factors.

Students may also encounter Newton's Law of Universal Gravitation, an inverse square law,

$$F = \frac{Gm_1m_2}{r^2}, \quad (2)$$

where $F$ is the force of gravity, $G$ is the gravitational constant, $m_1$ and $m_2$ are the masses of the two objects, and $r$ is the distance between the centers of mass of two objects. When applying equation 2, students may be asked to examine the direct relationship between $F$ and $m$, or the inverse square relationship between $F$ and $r$. Inverse square laws are seen multiple times in an astronomy course, for example, in the electromagnetic spectrum. When discussing light as a photon, the inverse square law is used to describe that intensity lessens with the square of the distance away from the source.

In terms of indirect relationships, students may calculate the temperature of a blackbody using Wien's Law:

$$\lambda_{max} = \frac{3 \times 10^6 \text{ nm·K}}{T}. \quad (3)$$

Here students examine the inverse relationship between the absolute temperature of an object and its peak wavelength, $\lambda_{max}$.

These mathematical relationships are indicative of how students might apply basic mathematical computation and understanding in introductory astronomy.

### C. Mathematics as a predictor of post-secondary STEM performance

Developmental mathematics coursework has often been identified as a roadblock for success in various measures of community college performance. Only 42% of high school students graduate with the skills necessary for college-level mathematics coursetaking [24]. Out of the remaining 58% of students who need to register for community college developmental mathematics courses, only one-third complete them [18]. Consequently, these students may not be able to register for the prerequisite college-level mathematics courses needed to enroll in STEM courses. This is also an equity issue since many students enrolling in developmental mathematics courses are low socioeconomic status students and/or underrepresented minorities in STEM [25,26].

Enrollment in developmental mathematics may discourage community college students from transfer and pursuing or completing a STEM degree [12,14,27-29]. Some students may delay taking required developmental mathematics courses which further increases time to degree completion or transfer [30]. However, research has found that students who completed the entire developmental mathematics sequence had similar outcomes to those students who entered at college-level mathematics [31].

### D. Theoretical Framework

The theoretical framework for the present study is based upon Bahr's deconstructive analysis approach for predicting community college students' academic outcomes [21]. This framework posits that students' pathways through higher education provide insights into their dynamic interaction with progress towards academic goals. To accomplish this, transcript data are used to "'deconstruct' the varied steps or stages through which students pass from the point of college entry to a given outcome of interest" [21, p.145]. Students' academic histories are examined through stages to identify course attempts and outcomes. In doing so, potential institutional interventions may be identified to improve performance and reduce attrition [21].

The three-fold mission of community colleges has been identified as: (1) upward transfer, (2) workforce development, and (3) community education [21]. Academic success is required for advancing towards a four-year degree, meeting readiness benchmarks for a desired career, and acquiring a credential. Transcripts, which provide road maps for academic decision making and performance, are useful in deconstructive analysis for understanding how students are academically integrated with higher education institutions [29]. Developmental mathematics courses are of specific interest in understanding successive academic competencies, since more than one-half of community college students enroll in these courses [20]. Upward transfer and workforce readiness are often predicated upon success in this coursework [21].

Although past research has often focused on demographic predictors of community college student outcomes [32,33], coursetaking and prior academic competencies may provide more useful insights into targeted interventions that influence student behaviors. This is particularly important in STEM disciplines since these careers have been identified as pathways to social and economic mobility for traditionally underserved students [34]. The deconstructive analytical approach is particularly useful in understanding how remedial pathways may be related to academic outcomes such as milestones, credential attainment, and transfer to four-year institutions [21].

The explanatory power of deconstructive analysis may provide insights on the diversification of STEM fields, which has been identified as a national priority in producing a STEM-literate citizenry and maximizing the nation's STEM enterprise [35]. Community colleges serve as a large talent pool for recruitment into science and technological careers [2].

## III. METHODS

### A. Context and data collection

This study is an observational, non-experimental study [36] that used longitudinal institutional data from a large public suburban New York State community college from the academic years 2015-2019. The use of de-identified data was approved by Stony Brook University's Institutional Review Board (#2021-00228). This community college had an approximate enrollment of 27,000 students during this time frame. The student body was 46.8% male and 53.2% female; 51.8% of the students were enrolled full-time. In terms of ethnicity, 54.7% were White, 23.8% Hispanic, 3.8% Asian/Pacific Islander, 8.0% African American, 0.3% Native American, and 9.5% other. Approximately 32.0% of the student population were ethnic minorities traditionally underrepresented in STEM, which included Hispanic, African American, and Native American students (as defined by the National Science Board [2]).

There were 2194 students who took astronomy during the 2015-2019 academic years. Students with missing information or a cumulative GPA or mathematics GPA of zero were disregarded. After filtering the missing information, the final data set consisted of transcripts for 1909 students.

At the community college where this study took place, students often enrolled in an astronomy course to fulfill their science general education requirement. The prerequisite for an astronomy course at this institution was Algebra I. Astronomy students were generally required to use basic algebra with scientific notation to solve simple astronomical problems as well as to understand conceptual applications of mathematical relationships. There were three different astronomy courses, which were intended to have the same level of difficulty and were stand-alone courses. Students could enroll in any number of these courses in any order. AST 1 addressed topics related to the solar system and the motion of celestial bodies in the night sky, AST 2 focused on topics related to stars, galaxies, and cosmology, and AST 3 discussed topics related to the search for life in the universe. Table I indicates topics covered in the three astronomy courses offered at the institution.

TABLE I. Description of prerequisites and topics addressed in astronomy courses

| Astronomy Course | Topics |
|---|---|
| AST 1 | 1. Historical Astronomy<br>2. Motions of Celestial Bodies<br>3. Naked Eye Astronomy<br>4. Planets and Moons of the Solar System<br>5. Asteroids and Comets |
| AST 2 | 1. Electromagnetic Spectrum<br>2. Stellar Classification and Evolution<br>3. Galaxies<br>4. Interstellar Medium<br>5. Modern Cosmology |
| AST 3 | 1. Scale of the Universe<br>2. Formation and Evolution of Life<br>3. Potential for Life in the Solar System<br>4. Exoplanet Detection<br>5. Interstellar Travel<br>6. Search for Extraterrestrial Intelligence |

## B. Statistical analyses

Presently, there is little or no research regarding predictors of performance in community college astronomy courses. This study examined demographic and academic variables, including gender, ethnicity, socioeconomic status, cumulative GPA, and mathematics coursetaking to predict student performance in community college astronomy. Statistical approaches were employed to answer the research question: How do students' academic coursework and performance and demographic characteristics predict introductory community college astronomy performance?

Logistic regression quantifies the odds of a particular variable's influence on the outcome and has been cited as one of the better models for making predictions and classifications [37]. This method reports probabilities of outcomes and does not assume normal distributions of predictor variables or errors. The binary academic outcome measured in this study was whether individual students passed one or more of the three astronomy courses with a grade of C or better.

## C. Operational variables

The researchers analyzed both academic and demographic variables as predictors of astronomy performance. Demographic variables included a student's ethnicity, gender, age, and socioeconomic status. Demographic variables were included in the model because prior research indicated demographics are often relevant predictors of a community college student's STEM outcomes [38-40]. Academic variables included students' grades, major (STEM or non-STEM), mathematics and science courses taken, academic grades in these courses, and full-time vs. part-time status. The number of astronomy classes taken was also included, as research has indicated students who have previously taken science classes in college have a better foundation of science concepts [41]. Variables were either continuous or dichotomous. Dichotomous variables were coded as either 0 or 1. A full description of the variable description, types, and coding can be seen in Table II.

Table II. Academic and demographic operational variables

| Variable Name | Description | Variable Category | Variable Type |
|---|---|---|---|
| **Cumulative GPA** | The student's cumulative grade point average on a four-point scale. | Academic | Continuous |
| **Enrolled in developmental mathematics** | Developmental mathematics courses included: Pre-Algebra, Algebra I, Developmental Mathematics Skills, Mathematical Literacy. Students who registered for any of these courses were grouped in the developmental mathematics category (yes = 1, no = 0). | Academic | Dichotomous |
| **Repeated mathematics courses** | A student's transcript revealed whether they repeated any mathematics course (mathematics repetition = 1, no repetition = 0). | Academic | Dichotomous |
| **Enrollment status** | Whether a student enrolled as a full-time or a part-time student (1 = full-time, 0 = part-time). | Academic | Dichotomous |
| **Major** | (STEM major = 1, non-STEM major = 0). | Academic | Dichotomous |
| **Ethnicity** | The ethnicity of a student and whether the student was an underrepresented minority in STEM. Consistent with classifications in federal reporting [2], underrepresented minorities in STEM included Black or African American, Hispanic, and Native Americans or Alaska | Demographic | Dichotomous |

| | Natives. Non-underrepresented minorities in STEM included White, Asian, and students who reported two or more ethnicities (URMS = 1, non-URMS = 0). | | |
|---|---|---|---|
| **Number of science courses taken** | The number of science courses in which students enrolled other than astronomy. | Academic | Continuous |
| **Number of mathematics courses taken** | The number of mathematics courses in which students enrolled. | Academic | Continuous |
| **Number of astronomy courses taken** | The number of astronomy classes in which a student enrolled, whether AST 1, AST 2, and/or AST 3. | Academic | Continuous |
| **Science performance** | Science course GPA not including astronomy courses on a four-point scale. | Academic | Continuous |
| **Mathematics performance** | Mathematics course GPA on a four-point scale. | Academic | Continuous |
| **Gender** | The self-reported gender of the student in binary terms (men = 1, women = 0) | Demographic | Dichotomous |
| **Socioeconomic status** | Socioeconomic status was determined by whether a student qualified for a Pell Grant based on their Expected Family Contribution (yes = 1, no = 0). | Demographic | Dichotomous |
| **Age** | Biological age in years. | Demographic | Continuous |
| **Astronomy performance** | The student's mean final grade in their astronomy courses. The grades awarded to students at this intuition were A, B+, B, C+, C, D+, D, F, and W (ABC = 1, DFW = 0). | Dependent | Dichotomous |

## IV. RESULTS

### A. Descriptive statistics

Descriptive statistics were generated to characterize the profiles of community college students taking astronomy and their academic performance. The average age of the student sample was 20.0 years old. Additionally, 58.1% of the astronomy students were men, 26.0% qualified for a Pell Grant, and 80.6% were not underrepresented minorities in STEM. The data indicated that 96.0% of students were non-STEM majors, indicating most STEM majors did not take astronomy.

In terms of mathematics remediation and performance, 28.0% of the students were required to register for some type of developmental mathematics course and 20.1% had to repeat at least one mathematics course. The average cumulative GPA for all astronomy students was 2.81, which is on cusp of the upper bound of the C+ grade. The mean mathematics GPA was 2.80, consistent with overall GPA.

Tables III and IV contain descriptive statistics of the academic data when comparing STEM and non-STEM majors in astronomy courses. The academic descriptive statistics are grouped by STEM and non-STEM majors. The choice to represent the data by major was to illustrate that even though astronomy was taken by STEM and non-STEM students, the majority of students enrolled in astronomy were non-STEM majors (consistent with [6]). Relatively few STEM majors took remedial mathematics, although their cumulative GPA was nearly identical to non-STEM majors.

TABLE III. *Academic descriptive statistics grouped by major*

| Major | Cumulative GPA | Took Developmental Mathematics | Repeated Math | Total |
|---|---|---|---|---|
| STEM | 2.80 ± 0.94 | 5 (6.5%) | 20 (26.3%) | 76 |
| Non-STEM | 2.80 ± 0.90 | 527 (28.8%) | 378 (20.6%) | 1833 |

TABLE IV. *Demographic descriptive statistics grouped by major*

| Major | Men | Women | URMS | Non-URMS | Qualified for Pell Grant | Age (years) | Total |
|---|---|---|---|---|---|---|---|
| STEM | 56 (73.7%) | 20 (26.3%) | 43 (56.6%) | 33 (33.4%) | 20 (26.3%) | 20.0 ± 3.0 | 76 |
| Non-STEM | 1053 (57.4%) | 780 (42.6%) | 328 (17.9%) | 1505 (82.1%) | 477 (26.0%) | 20.0 ± 3.0 | 1833 |

In terms of student enrollment in astronomy, 1762 students took only one astronomy course, 122 students took two astronomy courses, and 22 students took three astronomy courses. The performance in each of the three astronomy courses is represented in Table V, while the grade distributions are shown in Figure 1.

TABLE V. *Enrollment and grade distributions in each astronomy course*

|  | ABC | DFW | Total |
|---|---|---|---|
| AST 1 | 927 (70.1%) | 395 (29.9%) | 1,322 |
| AST 2 | 468 (74.9%) | 157 (25.1%) | 625 |
| AST 3 | 87 (79.8%) | 22 (20.2%) | 109 |

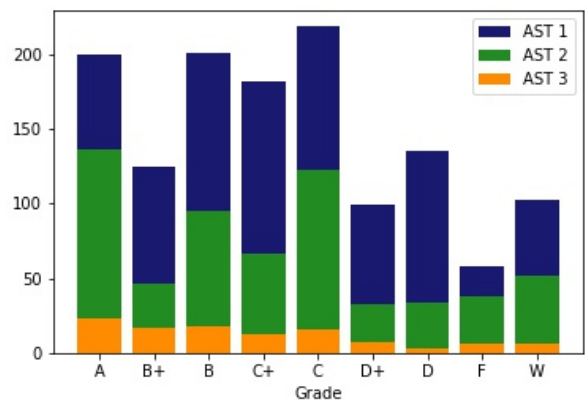

FIG. 1. *Grade distributions for astronomy courses*

## B. Independent variable selection

To determine significant predictors of astronomy performance, Spearman correlation coefficients were determined, and significant variables ($p<0.05$) were included in the multivariable model. Table VI indicates the correlation coefficients and $p$-values for the potential predictor variables and astronomy performance. None of the demographic variables were significant predictors of astronomy performance. The significantly correlated academic predictors were: (1) number of astronomy courses taken, (2) number of science courses taken other than astronomy, (3) STEM major, (4) whether a student took developmental mathematics, (5) whether a student repeated mathematics, (6) mathematics GPA, and (7) cumulative GPA. The variance inflation factors (VIF) were calculated for these selected values to identify potential multicollinearity, which indicates correlation between predictors and may confound the model. Cumulative GPA and mathematics GPA exceeded the tolerance threshold of VIF=5 [42]. Since cumulative GPA had a higher VIF and was partially calculated from mathematics GPA, it was removed from the multivariable model.

TABLE VI. *Correlations of demographic and academic variables with astronomy performance*

| Independent Variables | Correlation Coefficient ($r_S$) | $p$-value |
|---|---|---|
| **DEMOGRAPHIC VARIABLES** | | |
| Gender | 0.012 | 0.613 |
| Ethnicity | 0.000 | 0.993 |
| Socioeconomic status | 0.002 | 0.915 |
| Age | –0.005 | 0.841 |
| **ACADEMIC VARIABLES** | | |
| Number of astronomy courses taken* | 0.055 | 0.016 |
| Number of science courses taken other than astronomy** | –0.072 | 0.002 |
| Number of mathematics courses taken | –0.042 | 0.068 |
| Major (STEM or non-STEM)** | 0.064 | 0.005 |
| Enrolled in developmental mathematics*** | –0.158 | <0.001 |
| Repeated mathematics course(s)*** | –0.179 | <0.001 |
| Enrollment status (full-time/part-time) | 0.033 | 0.148 |
| Mathematics performance*** | 0.225 | <0.001 |
| Science performance in courses other than astronomy | –0.025 | 0.280 |
| Cumulative GPA*** | 0.329 | <0.001 |

*$p<0.05$; **$p<0.01$; ***$p<.001$

## C. Logistic regression model

A logistic regression model was generated to identify predictors of performance and the relationship of these predictors with the likelihood that students passed astronomy courses with a grade of C or better. The null hypothesis stated the categorical performance outcome was not statistically related to the significantly correlated academic predictors. The Wald chi-square statistic is the squared ratio of the estimate ($\beta$) to the standard error ($SE$), indicating the independent variable's predictive value in the model when considered with the associated $p$-value.

The model indicated four significant academic predictors of astronomy performance, including: (1) number of astronomy courses taken, (2) mathematics grade performance, (3) whether a student repeated mathematics, and (4) whether a student enrolled in developmental mathematics. The Pearson chi-square value ($\chi^2$=144.229, $p$<0.001) indicated the logistic model was significant, correctly classifying 71.2% of the cases, with a Nagelkerke $R^2$ of 0.104, a medium effect size. The Hosmer and Lemshow test indicated the model was a good fit ($\chi^2$=12.117, $p$=0.146).

For every astronomy class a student took, they were 71% more likely to pass the course. For every increased grade point earned in a mathematics course, students were 54% more likely to pass their astronomy course. If students repeated their mathematics courses, they were 33% more likely to fall in the DFW category in astronomy than those who did not repeat mathematics courses. Students enrolling in a developmental mathematics course were 43% less likely to pass astronomy than those who did not take developmental mathematics. Notably, STEM major and the number of science classes taken did not predict astronomy performance in the multivariable model. The logistic regression statistics are summarized in Table VII.

TABLE VII. *Logistic regression model for astronomy performance*

| Independent Variables | $\beta$ (*SE*) | Wald | Odds Ratio Exp ($\beta$) | 95% Confidence Interval for Odds Ratio | *p*-value |
|---|---|---|---|---|---|
| Major | 0.705 (0.365) | 3.725 | 2.024 | [0.989, 4.142] | 0.054 |
| Number of astronomy courses taken** | 0.534 (0.201) | 7.036 | 1.705 | [1.150, 2.530] | 0.008 |
| Mathematics performance*** | 0.432 (0.068) | 40.726 | 1.540 | [1.349, 1.759] | <0.001 |
| Number of science courses taken | −0.052 (0.043) | 1.449 | 0.950 | [0.873, 1.033] | 0.229 |
| Repeated mathematics courses** | −0.400 (0.137) | 8.487 | 0.671 | [0.513, 0.877] | 0.004 |
| Enrolled in developmental mathematics*** | −0.565 (0.113) | 25.125 | 0.568 | [0.456, 0.709] | <0.001 |

**$p$<0.01; ***$p$<.001

## V. DISCUSSION

The results from this study are useful in understanding how community college students, a large talent source for STEM careers, might improve astronomy performance and science literacy. The first major finding of this study was that demographic variables were not significant predictors of astronomy performance, which contradicts prior research on student outcomes in community college settings [18-20]. For example, Bailey et al. [18] and Maxwell et al. [20] found that women, Black, and Hispanic community college students tended to enroll in developmental coursework at disproportionately high rates. However, Black students were less likely to progress through the sequence than White students while women were more likely to progress through the sequence than men (there was no difference for Hispanic students) [18]. Bahr et al. [43] also found that women were more likely to persist towards a community college credential than men. Another study indicated that the proportion of minority student population in community colleges was negatively related to the likelihood of graduation or transfer [19]. These studies varied in their explanations for findings related to disparate impacts, for example, suggesting marginalizing climates for students traditionally underrepresented in STEM [43], limited opportunities to form learning communities [20], burnout for those with lower-level entry points [43], and lack of academic preparation [19].

The present study aligns with other research that reported demographic variables are not significant predictors of performance in STEM course performance, graduation, and transfer in college settings [14,17,44,45]. Community colleges are open enrollment institutions, and 40% of students enrolled nationally are underrepresented minorities in STEM [3]. Research has shown students who complete their developmental mathematics courses have equivalent academic success when compared to students who do not require developmental mathematics [31,46]. Since developmental mathematics completion was the sole prerequisite for the astronomy courses in this study (and presumably in other community colleges), more students have the opportunity to select this science to complete curricular requirements. Consequently, the results from this study suggest introductory community college astronomy may be an equalizer for students of different ethnic and socioeconomic backgrounds. Enrolling in a community college astronomy course has the potential to make STEM fields accessible to students who may otherwise might not have this aspiration or opportunity.

The second major finding was the role of mathematics in predicting astronomy performance. Students who took developmental mathematics, repeated mathematics, and did not perform well in mathematics were more likely to receive a D, F, or W in their astronomy classes. This finding was consistent with research that identified developmental mathematics coursetaking as a predictor of community college science performance and STEM attrition [12,14,16,27,47], however, these studies did not examine astronomy specifically.

Deconstructive analysis [21,31,46] suggests that understanding community college students' academic pathways may provide insights on how their outcomes might be improved. Mathematics has been shown to influence STEM pathways starting at the precollege level, where performance metrics are often used to identify students for post-secondary remediation. At the community college level, failure and repetition in mathematics coursework may increase financial burdens, graduation delays, disinterest in STEM disciplines, STEM attrition, and social stratification [12,14,29,46,48]. This has been attributed to mathematical skill deficiency, delayed enrollment in remedial mathematics, and inadequate college advisement [29,46]. The results of this study suggest that poor performance in mathematics has a ripple effect on performance in other STEM disciplines – in this case, astronomy. The potential of astronomy as a gateway to STEM may not be fully realized without addressing student achievement in mathematics at multiple points in the STEM pipeline.

An additional finding was that students who enrolled in multiple astronomy courses tend to earn higher grades in these courses. This is a more complex construct to explain since the direction of potential causality is ambiguous. Students who took more than one astronomy course may have done so for several reasons. Students' interest may have correlated to their initial performance in astronomy; research has shown that high performance and conceptual knowledge may reinforce academic self-concept in science [48]. This identity may have led to enrollment in other astronomy courses. Conversely, students may have first developed an interest in astronomy, devoted more effort in their coursework, and chose subsequent enrollment due to this interest. In either case, positive experiences in introductory astronomy and physics coursework may be a key predictor of STEM enrollment, interest, and related career intentions [49].

## VI. IMPLICATIONS

The results from the present study suggest several implications for research and practice. Studying academic preparation, coursetaking, and performance helps shed light on how enrollment in developmental mathematics coursework might be revisited to improve the outcomes of community college astronomy students. This coursework, along with precollege preparation in mathematics, may hinder students from progressing academically, delaying or restricting graduation, upward transfer, and four-year college degree completion. The findings of this study may inform interventions regarding mathematics coursetaking and performance to improve STEM performance and persistence for both STEM and non-STEM students. Recommendations are discussed related to precollege academic preparation, astronomy outreach, and academic counseling.

Precollege mathematical preparation may influence student astronomy outcomes in higher education. Mathematical remediation and college bridge programs have been shown to improve STEM outcomes for students with inadequate mathematical opportunities [16,31]. Mathematics is a critical gatekeeper in STEM persistence, and its role in astronomy performance should be given careful consideration throughout the pipeline [13].

Regarding astronomy interest, high school astronomy coursework or public outreach events may promote positive attitudes before students enroll in community colleges [50]. Engagement with informal science institutions has the potential to spark student's interest to pursue astronomy earlier in their academic pathways [51]. Research has shown that interest in astronomy in high school may be a feeder for students to study STEM fields in college [50].

Formal advisement for community college students is another recommendation for improving astronomy outcomes. Students may benefit from academic advisement to enroll in developmental mathematics in their first semester and subsequently continue with astronomy coursework. Research has shown that early enrollment in developmental mathematics improves academic performance and persistence [18]. Some community college students may not be socially integrated with the institution and thus unable to make informed decisions about their academic pathways [52]. Without adequate advisement from "institutional agents," students may misjudge their need for remedial instruction and diminish their academic commitment, resulting in persistent social stratification [29,52]. This advisement may be particularly beneficial for students who intend to choose astronomy to fulfill science requirements, since they may not be aware of the fundamental mathematical applications.

Future research on community college STEM courses should focus on institutional factors rather than demographics, since it may be more effective to strategize on improving academic predictors that may affect astronomy performance. An example of institutional focus is the developmental mathematics sequence and how that affects student performance in other STEM courses.

## VII. LIMITATIONS

This study explored predictors of community college astronomy course performance. The data used in this study were localized in New York State and did not include student data from other geographic regions. Therefore, the findings may not be generalizable to all community college students. The transcript analysis did not include consideration of course sequences and whether students received advisement on academic pathways. Data were not collected for students who continued enrollment beyond 2019, which may have limited the analysis of longitudinal academic patterns. Furthermore, pedagogical practices among astronomy faculty were not examined; differences in classroom climate and instruction of mathematical astronomy concepts may have skewed the results. Since quantitative analyses may have limited explanatory power, qualitative research with a purposeful sample of diverse community college astronomy students may reveal insights on students' STEM-related academic and career intentions.

## VIII. CONCLUSIONS

The results from this study revealed that student performance in community college astronomy courses was predicted by mathematics performance and taking multiple astronomy courses, suggesting that institutions may design interventions that target students with mathematical challenges to improve astronomy outcomes. Ethnicity, gender, age, and socioeconomic status did not predict performance, which contradicted some prior research in higher education STEM outcomes [18-20]. These results imply that astronomy coursetaking may promote equity in science participation and performance. Community college policy makers may consider revisiting mathematical remediation and academic advising to promote student achievement in STEM coursework. These findings also suggest further study is required regarding the role of astronomy enrollment and performance in promoting interest in STEM, choice of STEM major, and scientific literacy.

## ACKNOWLEDGMENTS

The authors would like to thank Richard Weiner for his assistance with database compilation. Thanks also to the anonymous reviewers for their insightful recommendations to improve the manuscript.